\documentclass[aps,amsfonts,nofootinbib,superscriptaddress,preprint]{revtex4-1}
\usepackage{comment}
\usepackage[normalem]{ulem} 
\makeatletter
\usepackage{amsmath}
\usepackage{amssymb}
\usepackage{appendix}
\usepackage{amsbsy}
\usepackage{bm}
\usepackage{epsfig} 
\usepackage{color}
\usepackage{slashed}
\usepackage{soul}
\usepackage{comment}
\makeatletter
\usepackage{amsmath}
\usepackage{amssymb}
\usepackage{appendix}
\usepackage{amsbsy}
\usepackage{bm}
\usepackage{epsfig} 
\usepackage{color}
\usepackage{slashed}
\newcommand{\stkout}[1]{\ifmmode\text{\sout{\ensuremath{#1}}}\else\sout{#1}\fi}

\newcommand{\bb}{\begin{equation}}

\newcommand{\ee}{\end{equation}}

\newcommand{\eqb}{\begin{eqnarray}}

\newcommand{\eqf}{\end{eqnarray}}

\def\pivec{\mbox{\boldmath$\pi$}}
\def\sigmavec{\mbox{\boldmath$\sigma$}}

\usepackage[T1]{fontenc}

\begin{document}
\title{  Relativistic Topological Insulator Model }

\author{J. Gamboa}
\email{jorge.gamboa@usach.cl}
\affiliation{Departamento de F\'isica, Universidad de Santiago de Chile, Casilla 307, Santiago, Chile}
\author{F. M\'endez}
\email{fernando.mendez@usach.cl}
\affiliation{Departamento de F\'isica, Universidad de Santiago de Chile, Casilla 307, Santiago, Chile}

\begin{abstract}

A relativistic topological insulator model in three spatial dimensions which is a non 
trivial extension  of the non-abelian Landau problem is proposed. The model is exactly soluble and energy levels have both a discrete  and a continuous  degeneracy.  The chromomagnetic field is strong and the fermions are confined in a plane and  the physical effects that appear reflects the ${\bm Z}_2$ symmetry.
\end{abstract}
\maketitle

 The physics of a topological insulator is an interesting field of research not only 
 because of its interest in condensed matter physics and statistical mechanics but 
 also because of its potential applications in other fields of physics \cite{kane1x}.

In general, a topological insulator resembles the FQHE but does not require an 
external magnetic field which, formally, is hidden in matter by a topological 
protection mechanism \cite{moore}. The technical way to implement these properties is 
due to Kane and Mele who proposed to add a spin-orbit term to the Hamiltonian in order 
to impose reversal-time invariance \cite{kane1,fu}.

In $2 + 1$-dimensions this procedure was also studied by Benevig and Zhang who 
explicitly showed how a topological insulator mimics as two Landau problems with 
opposite magnetic fields thus providing the $Z_2$-symmetry \cite{zhang}.

Under these conditions it seems natural to ask how this viewpoint works in the 
relativistic domain and how these ideas could provide a new approach to problems in 
high energy physics and mathematical physics \cite{nos}.

In this letter we would like to propose an example of a relativistic system that 
contains all the properties of a topological insulator, {{is also exactly soluble and it can be the starting point for modeling dark matter problems. }}

Although the model we propose looks like Moshinsky's Dirac oscillator \cite{moshinsky} 
from both a physical and a mathematical point of view, it is very different. Some 
properties that the model contains are, on the one hand, a closed connection with the 
QHE and non-commutative geometry and on the other, {{it contains all the physical properties of a true topological insulator.}}

In order to formulate the model let us consider the  non-trivial modification of the 
Dirac oscillator where, instead of making the substitution $ p_i \to \pi_i= p_i + i m 
\beta \omega x_i $ as in \cite {moshinsky}, we  analyze 
$$
i \frac{\partial \psi}{\partial t} = H \psi,
$$
for the Hamiltonian
\bb
\label{ham1}
H= \left( 
\begin{array}{lr} 
m  & \sigmavec\cdot\pivec 
\\
 \sigmavec\cdot\pivec  &-m
 \end{array} 
 \right), 
 \ee
with $\pivec$ defined as 
\bb
\pi_i = \left\{
\begin{array}{lll}  \left(
p_i +  B \epsilon_{ij} x_j\otimes \sigma_3\right), &   &\{i,j\}=1,2, \label{pi}
 \\
p_3, &    & ~~~~~~i=3.
\end{array}
\right.
\ee
Here $B$ is a constant with canonical dimension $+2$ and $\sigma_3$ is defined  in the 
color space. Note that
the products with identity has been omitted, as for example $m\otimes \openone$ and 
$p_3\otimes\openone$.

Note that the operators $x_i$ and $\pi_i$ satisfy 
\eqb
\left[x_i,\pi_j\right] &=& i \delta_{ij},
 \nonumber 
\\
\left[\pi_i,\pi_j\right] &=&2 i B \epsilon_{ij} \sigma_3, \label{ss}
\eqf
and zero in all other cases.

The quantity $A^a_i = \epsilon_{ij}Bx_j\,\sigma^a$ with $a\in\{1,2,3\}$ is an element 
of  the $SU(2)$ algebra, and then, 
the commutator (\ref{ss}) indicates that in the RHS,  $A^3_i=\epsilon_{ij} B x_j 
\sigma_3$ is a component of the $SU(2)$ gauge potential
in the (internal )direction $a = 3$ equivalent to   a constant chromomagnetic field 
\cite{weis}.  Relation (\ref{ss}) furnishes  an example 
of a deformed commutator as those appearing in non commutative quantum 
field theory,  graphene or  in non-commutative geometry.

 The Hamiltonian in (\ref{ham1}) can be written as follows 
\begin{equation}
H= \left(
\begin{array}{cc}
m\,\openone\otimes\openone & \boldsymbol{\sigma}\cdot{\bf p}\otimes\openone
+B(\boldsymbol{\sigma}\times{\bf x})\cdot\hat{z}\otimes\sigma_3
\\
\boldsymbol{\sigma}\cdot{\bf p}\otimes\openone
+B(\boldsymbol{\sigma}\times{\bf x})\cdot\hat{z}\otimes\sigma_3 & -m\,
\openone\otimes\openone
\end{array}\right),
\end{equation}

Since $H$ is independent of time   we look for solutions with  the form
\bb
\label{solfi}
\psi ({\bf x},t)= e^{-i E\,t} 
\left( 
\begin{array}{l} 
\Phi_1({\bf x}) 
\\ \Phi_2({\bf x}) 
\end{array} 
\right),
\ee
with
\begin{equation}
\label{soll}
\Phi_1=\phi_1\otimes \left(
\begin{array}{c}
1
\\
0
\end{array}\right) + 
\phi_2 \otimes \left(
\begin{array}{c}
0
\\
1
\end{array}\right),
\quad
\Phi_2=\chi_1\otimes \left(
\begin{array}{c}
1
\\
0
\end{array}\right)
+
\chi_2\otimes \left(
\begin{array}{c}
0
\\
1
\end{array}\right),
\end{equation}
and $\{\phi_i,\chi_j\}_{\{i,j\}\in\{1,2\}}$, two-components spinors.

Then the Dirac equation reads 
\begin{eqnarray}
\label{osc33}
\bigg[ \boldsymbol{\sigma}\cdot{\bf p}\otimes\openone
+B(\boldsymbol{\sigma}\times{\bf x})\cdot\hat{z}\otimes\sigma_3\bigg]\Phi_2 
&=&(E-m)\openone\otimes \openone\, \Phi_1  ,
\\
\label{osc34}
\bigg[ \boldsymbol{\sigma}\cdot{\bf p}\otimes\openone
+B(\boldsymbol{\sigma}\times{\bf x})\cdot\hat{z}\otimes\sigma_3\bigg]\Phi_1
&=& (E+m)\openone\otimes \openone\, \Phi_2.
\end{eqnarray}
This equation can be decoupled by using standard methods (see supplementary material in the appendix) to
get
\begin{equation}
\label{osc35}
\bigg[\big(
{\bf p}_\perp^2+p_3^2 + B^2{\bf x}_\perp^2
\big)\otimes\openone -2B(L_3+\sigma_3)\otimes\sigma_3\bigg]\Phi_1
=
(E^2-m^2)\openone\otimes\openone\,\Phi_1,
\end{equation}
and similar expression for $\Phi_2$. Here where the index  $_\perp$ denotes  
quantities in the plane $x_1-x_2$  and the  angular momentum along the axis $x_3$ is  
$L_3=x_1p_2-x_2p_1$.
From now on, we will focus only in the solution for $\Phi_1$.

%

The term $-2 B L_3 \sigma_3$ is the spin orbit coupling containing the up and down 
projections of the magnetic field and $B\sigma_3 $  is a constant that in the end will 
only contribute to the energy spectrum.

Operator in (\ref{osc35}) is  diagonal and commutes with $p_3\otimes\openone$. Therefore
we look for solutions with the form
$$
\Phi_1=e^{i p_zx_3}
\left(
\begin{array}{c}
\varphi_1(x_1,x_2)
\\
\varphi_2(x_1,x_2)
\end{array}
\right).
$$
Equivalently $\phi_i = e^{\imath\,p_z\,x_3}\varphi_i$ in (\ref{soll}), and then
 spinors  $\varphi_i$ satisfy
\begin{eqnarray}
\label{ec1}
\left({\bf p}_\perp^2+ B^2{\bf x}_\perp^2-2B(L_3+\sigma_3) \right)\varphi_1 &=& 
(E^2-m^2-p_z^2)\varphi_1,
\\
\label{ec2}
\left({\bf p}_\perp^2+ B^2{\bf x}_\perp^2+2B(L_3+\sigma_3) \right)\varphi_2 &=& 
(E^2-m^2-p_z^2)\varphi_2,
\end{eqnarray}

%
%

Equations (\ref{ec1}) and (\ref{ec2})  are  related through the transformation
 $B \to -B$. Therefore, the two independent equations are
\bb
\left({\bf p}_\perp^2 + B^2 {\bf x}_\perp^2 - 2 BL_3  \right)  \varphi_\pm = \left(E^2 -
m^2 -p_z^2 \pm 2B\right) \varphi_\pm.  \label{osc8} 
\ee
Here $\varphi_\pm$ denotes the two components of the spinor $\varphi_1$, and the 
two components of $\varphi_2$ once the $B\to -B$ transformation has been performed.

For convenience let us define (\ref{osc8}) in terms of the dimensionless variables
\[
\tilde{ {\bf x}}= \sqrt{B}\, {\bf x}_\perp, 
\]
so  (\ref{osc8})  becomes
\bb
\left({\tilde  {\bf p}}^2 + { \tilde {\bf x}}^2 - 2 L_3  \right) \varphi_\pm = \left(\frac{E^2 -m^2 - p_z^2}{B}\pm 2\right) \varphi_\pm.  \label{osc888} 
\ee
with $\tilde{{\bf p}}^2 =-\nabla_{\tilde{x}}^2$.


In order to find the explicit solutions, let us note that the effective movement takes 
place in the plane $x_1-x_2$ and it is enough to solve the equation (\ref{osc8}) in 
polar coordinates. 

We look for solutions with the form $\varphi_\pm  \sim R_\pm (r)\,e^{i\ell\theta}$, 
where  $\ell\in {\bm Z}$,  $\theta$ is the polar angle and $r^2=\tilde{x}_1^2 +
\tilde{x}_2^2$.   
The equation (\ref{osc8}) becomes
\begin{equation}
R_{\pm}''+\frac{R_{\pm}'}{r}+\left[{\cal E}_\pm^2-r^2-\frac{\ell^2}{r^2}\right]
\,R_\pm=0,
\end{equation}
with ${\cal E}_\pm^2=\frac{E^2 -m^2 -p_z^2}{B} + 2(\ell\pm 1) $. 

The change of variables 
\begin{equation}
R_\pm(r) \propto e^{-\frac{r^2}{2}}\,r^{|\ell|} g_\pm(r),
\end{equation}
yields  to 
\bb
g_\pm''(r) + \left( \frac{2|\ell| +1}{r} - 2 r\right)g_\pm'(r) + \left({\cal E}_\pm^2 - 2(|
\ell| +1) \right) g_\pm(r)=0. \label{last}
\ee

Finally making $\xi =r^2$ in (\ref{last}), we obtain the confluent hypergeometric 
equation 
\bb
\xi g_\pm''(\xi) + \left(|\ell| +1-\xi \right) g_\pm'(\xi) + \left( 
\frac{{\cal E}^2_{\pm}}{4}-\frac{|\ell|+1}{2} \right) g_\pm(\xi)=0, \label{conflu}
\ee
The requirement to have a square integrable solution   for  $\xi\in[0,\infty)$ imposes 
\bb
\frac{{\cal E}_{\pm}^2}{4}-\frac{|\ell| +1}{2}  = n_\pm, \label{sol2}
\ee
where $n_\pm \in \mathbb{Z}$, implying the spectrum
\bb
E^2_\pm =   m^2 + p_z^2 + 2B \left(2 n_\pm + |\ell |-\ell+1\pm 1 \right). 
\label{spectrum}
\ee

There is an apparent inconsistency in the spectrum (\ref{spectrum}), however since 
particle-antiparticle symmetry implies that $n_- \leftrightarrow n_+$ and $\ell 
\leftrightarrow -\ell$, the condition $E_-=E_+$ yields 
\bb
n_--n_+ = 1. \label{depe}
\ee

The last equation shows the complete symmetry of the particles and antiparticles 
spectrum.

The convergent solution of (\ref{conflu}) is the hypergeometric function $F(n_\pm, |
\ell| + 1,r^2)$ --except by a  normalization constant-- which coincides with the 
associated  Laguerre  polynomials $L^{|\ell|}_{n_\pm} (r^2)$.

Finally, restoring the original variables, the component $\Phi_1$ (see 
(\ref{solfi}) and (\ref{soll})) turn out to be
\begin{eqnarray}
\label{finfi}
({\Phi_1})^\ell_n({\bf x}) &=& e^{i\,p_z\,x_3 + \imath\,\ell\theta}
\,e^{\frac{B\rho^2}{2}}\,\rho^{|\ell|}
\bigg{[}
\left(
\begin{array}{c}
{\cal C}_+\, L_{n}^{|\ell|}(B \rho^2)
\\
{\cal C}_-\, L^{|\ell|}_{n+1} (B\rho^2)
\end{array}
\right) 
\otimes 
\left(
\begin{array}{c}
1
\\
0
\end{array}
\right)
+
\left(
\begin{array}{c}
{\cal D}_+\, L_{n}^{|\ell|}(B \rho^2)
\\
{\cal D}_-\,L^{|\ell|}_{n+1}(B\rho^2)
\end{array}
\right)
\otimes 
\left(
\begin{array}{c}
0
\\
1
\end{array}
\right)
\bigg]
\nonumber
\\
&&
\end{eqnarray}
where ${\cal C}_\pm$ and ${\cal D}_\pm$ are normalization constants and $\rho^2
=x_1^2+x_2^2$.

The other component, namely $\Phi_2$, has the same form since it is the solution of the
same operator appearing in (\ref{osc35}). However, it is not independent of $\Phi_1$ due
to the relation (\ref{osc34}). In other words, given $\Phi_1$ in (\ref{finfi}), the
second component $\Phi_2$ in (\ref{solfi}) is
$$
({\Phi_2})^\ell_n ({\bf x})= (E+m)^{-1}
\bigg[ \boldsymbol{\sigma}\cdot{\bf p}\otimes\openone
+
B(\boldsymbol{\sigma}\times{\bf x})\cdot\hat{z}\otimes\sigma_3\bigg](\Phi_1)^\ell_n
$$

It is worth mentioning that even though in the model discussed in this paper the 
chromomagnetic field is strong and the fermions are confined in a plane, the $\mp B $ 
effect that appears in (\ref{osc8}) reflects the ${\bm Z}_2$ symmetry of the model 
from Benevig-Zhang \cite{zhang}.

The model discussed in this paper can be extended in various directions such as spin 
non-commutativity \cite{spin}, graphene in the sense discussed in \cite{grafeno}. 
{{However, the most interesting idea is the potential applications to dark matter and the analogies that can be established with topological insulators. 

We will discuss these ideas in a forthcoming paper.}}
\section*{Acknowledgements}

It is my pleasure to thank Prof. A. P. Balachandran for the discussions and comments. 
 One of us (J.G.) thanks  the Alexander von Humboldt Foundation by  
 support.   This research was supported by DICYT 042131GR (J.G.) and 041931MF (F.M.).

\appendix
\section{Supplementary material}
Consider (\ref{ham1}) written in the form 
\begin{equation}
H= \left(
\begin{array}{cc}
m\,\openone\otimes\openone & \boldsymbol{\sigma}\cdot{\bf p}\otimes\openone
+B(\boldsymbol{\sigma}\times{\bf x})\cdot\hat{z}\otimes\sigma_3
\\
\boldsymbol{\sigma}\cdot{\bf p}\otimes\openone
+B(\boldsymbol{\sigma}\times{\bf x})\cdot\hat{z}\otimes\sigma_3 & -m\,
\openone\otimes\openone
\end{array}\right),
\end{equation}
with $\hat{z}$ the direction $i=3$ in space. We look for solutions with 
the form
$$
\Psi =e^{-\imath Et} \left(
\begin{array}{c}
\Phi_1({\bf x})
\\
\Phi_2({\bf x})
\end{array}
\right),
$$
and therefore, equation of motion reads 
\begin{eqnarray}
\bigg[ \boldsymbol{\sigma}\cdot{\bf p}\otimes\openone
+B(\boldsymbol{\sigma}\times{\bf x})\cdot\hat{z}\otimes\sigma_3\bigg]\Phi_2 
&=&(E-m)\openone\otimes \openone\, \Phi_1  ,
\\
\bigg[ \boldsymbol{\sigma}\cdot{\bf p}\otimes\openone
+B(\boldsymbol{\sigma}\times{\bf x})\cdot\hat{z}\otimes\sigma_3\bigg]\Phi_1
&=& (E+m)\openone\otimes \openone\, \Phi_2.
\end{eqnarray}
This system can be decoupled in the standard way. For example, if we 
multiply firs equation by $(E+m)\openone\otimes\openone$ we obtain
$$
\bigg[ \boldsymbol{\sigma}\cdot{\bf p}\otimes\openone
+B(\boldsymbol{\sigma}\times{\bf x})\cdot\hat{z}\otimes\sigma_3\bigg]
((E+m)\,\openone\otimes\openone)\Phi_2 
=(E^2-m^2)\,\openone\otimes \openone\, \Phi_1,
$$
and we can replace now $\Phi_2$ from the second equation to obtain 
\begin{equation}
\label{ap1}
\bigg[ \boldsymbol{\sigma}\cdot{\bf p}\otimes\openone
+B(\boldsymbol{\sigma}\times{\bf x})\cdot\hat{z}\otimes\sigma_3\bigg]^2
\, \Phi_1 
=(E^2-m^2)\,\openone\otimes \openone\, \Phi_1.
\end{equation}
Let us evaluate all terms in the LHS in previous expression.
\begin{equation}
\label{cuad1}
({\boldsymbol \sigma}\cdot{\bf p})={\bf p}^2,
\end{equation}
since $\sigma_i\sigma_j = \delta_{ij}+\imath\,\epsilon_{ijk}\sigma_k$.
On the other hand,
\begin{eqnarray}
\label{cuad2}
[({\boldsymbol \sigma}\times {\bf x})\cdot\hat{z}]^2 &=&
\epsilon_{ij3}\,\sigma_i\,x_j\,\epsilon_{mn3}\,\sigma_m\,x_n 
\nonumber
\\
&=&x_j\,x_n\,\epsilon_{ij3}\,\epsilon_{mn3}(\delta_{im}+\imath\,\epsilon_{imk}\sigma_k)
\nonumber
\\
&=&
x_j\,x_n(\epsilon_{ij3}\,\epsilon_{in3}+\imath\sigma_k \epsilon_{ij3}\epsilon_{mn3}
\epsilon_{imk})
\nonumber
\\
&=&x_j\,x_n(\delta_{jn}\delta_{33}-\delta_{j_3}\delta_{n3})
+x_j\,x_n\,\imath\,\sigma_k\,\epsilon_{mn3}(\delta_{jm}\delta_{3k}-\delta_{jk}
\delta_{3m})
\nonumber
\\
&=&{\bf x}^2-(x_3)^2+\imath(x_m\,x_n\,\epsilon_{mn3}\,\sigma_3 + 
{\bf x}\cdot{\boldsymbol \sigma}\,x_n\epsilon_{n33})
\nonumber
\\
&=&x_1^2+x_2^2
\nonumber
\\
&\equiv& {\bf x}_\perp^2.
\end{eqnarray}
The last term is (up to the constant factor $B$)
$$
({\boldsymbol \sigma}\cdot{\bf p})(( {\boldsymbol \sigma}\times{\bf x})\cdot\hat{z})+
( ({\boldsymbol \sigma}\times{\bf x})\cdot\hat{z})({\boldsymbol \sigma}\cdot{\bf p})
\equiv 
\{{\boldsymbol \sigma}\cdot{\bf p},({\boldsymbol \sigma}\times{\bf x})
\cdot\hat{z}\},
$$
Let us write both terms separately 
\begin{eqnarray}
({\boldsymbol \sigma}\cdot{\bf p})(( {\boldsymbol \sigma}\times{\bf x})\cdot\hat{z})&=&
(\sigma_1p_1+\sigma_2 p_2+\sigma_3 p_3)(\sigma_1x_2-\sigma_2 x_1)
\nonumber
\\&=&p_1x_2-p_2x_1-\sigma_1\sigma_2 p_1x_1+\sigma_2\sigma_1 p_2x_2+\sigma_3\sigma_1 p_3x_2
-\sigma_3\sigma_2 p_3x_1
\nonumber
\\
(( {\boldsymbol \sigma}\times{\bf x})\cdot\hat{\bf p})({\boldsymbol \sigma}\cdot{\bf p})
&=&
(\sigma_1x_2-\sigma_2x_1)(\sigma_1p_1+\sigma_2p_2+\sigma_3p_3)
\nonumber
\\&=&p_1x_2-p_2x_1 + \sigma_1\sigma_2 x_2p_2 - \sigma_2\sigma_1 x_1p_1 +\sigma_1\sigma_3 x_2p_3 - \sigma_2\sigma_3 x_1 p_3,
\end{eqnarray} 
therefore
\begin{eqnarray}
\{{\boldsymbol \sigma}\cdot{\bf p},({\boldsymbol \sigma}\times{\bf x})
\cdot\hat{z}\} &=&
-2L_3+\sigma_1\sigma_2(x_2p_2-p_1x_1)+\sigma_2\sigma_1(p_2x_2 - x_1p_1)
+
\nonumber
\\
&&p_3 x_2(\sigma_3\sigma_1 + \sigma_1 \sigma_3) - x_1 p_3 (\sigma_3\sigma_2 
+ \sigma_2\sigma_3) 
\nonumber
\\
&=& -2L_3 +\sigma_1\sigma_2(x_2p_2-p_2x_2+x_1p_1-p_1x_1),
\end{eqnarray}
where, in the last line, we have used $\sigma_1\sigma_2=-\sigma_2\sigma_1$ 
and $\sigma_3\sigma_2 = - \sigma_2\sigma_3$.
Finally,
\begin{eqnarray}
\label{cuad3}
\{{\boldsymbol \sigma}\cdot{\bf p},({\boldsymbol \sigma}\times{\bf x})
\cdot\hat{z}\} &=& -2L_3+\sigma_1\sigma_2([x_1,p_1]+[x_2,p_2])
\nonumber
\\
&=&-2L_3+2\imath\,\sigma_1\sigma_2
\nonumber
\\
&=& -2L_3  - 2\sigma_3
\end{eqnarray}

By replacing (\ref{cuad1}), (\ref{cuad2}) and (\ref{cuad3}) in (\ref{ap1})
we obtain the equation for $\Phi_1$
\begin{equation}
\bigg[\big(
{\bf p}_\perp^2+p_3^2 + B^2{\bf x}_\perp^2
\big)\otimes\openone -2B(L_3+\sigma_3)\otimes\sigma_3\bigg]\Phi_1
=
(E^2-m^2)\openone\otimes\openone\,\Phi_1
\end{equation}
where the subindex  $_\perp$ denotes  quantities in the plane $x_1-x_2$  and the 
angular momentum along the axis $x_3$ is  $L_3=x_1p_2-x_2p_1$.

Previous equation can be recast in matrix form as follows
\begin{eqnarray}
\left(
\begin{array}{cc}
E^2-m^2-p_3^2-{\bf p}_\perp^2-B^2{\bf x}_\perp^2 + 2B(L_3+\sigma_3) & 0
\\
0 & E^2-m^2-p_3^2-{\bf p}_\perp^2-B^2{\bf x}_\perp^2 - 2B(L_3+\sigma_3)
\end{array}
\right)\Phi_1 &=&0
\nonumber
\\
\end{eqnarray}

Since $p_3$ is a conserved quantity, we look for solutions with the form
$$
\Phi_1=e^{\imath p_zx_3}
\left(
\begin{array}{c}
\varphi_1(x_1,x_2)
\\
\varphi_2(x_1,x_2)
\end{array}
\right).
$$
The spinors  $\varphi_i$ satisfy
\begin{eqnarray}
\left({\bf p}_\perp^2+ B^2{\bf x}_\perp^2-2B(L_3+\sigma_3) \right)\varphi_1 &=& 
(E^2-m^2-p_z^2)\varphi_1,
\\
\left({\bf p}_\perp^2+ B^2{\bf x}_\perp^2+2B(L_3+\sigma_3) \right)\varphi_2 &=& 
(E^2-m^2-p_z^2)\varphi_2,
\end{eqnarray}

\end{document}